\documentclass[fleqn,usenatbib, openany]{mnras}
\usepackage{amsmath}
\usepackage{textcomp}
\usepackage{gensymb}
\usepackage[hang,flushmargin,stable]{footmisc} 
\usepackage{graphicx}
\usepackage{txfonts}
\usepackage{xcolor}
\usepackage{float}
\usepackage{hyperref}
\usepackage{makecell}
\usepackage{diagbox}
\usepackage{threeparttable}
\newcommand{\orcid}[1]{\href{https://orcid.org/#1}{\includegraphics[width=8pt]{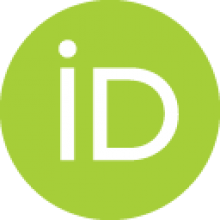}}}

\graphicspath{{./}{Figures_20220513}}

\title[Asymmetrical Mass Ejection from Proto-White Dwarfs]{Impact of Asymmetrical Mass Ejection from Proto-White Dwarfs on the Properties of Binary Millisecond Pulsars}

\author[Tang, Gao, \& Li]{
	Wen-Shi Tang$^{1,~2}$,
	Shi-Jie Gao$^{1,~2}$,
	Xiang-Dong Li$^{1,~2}$\thanks{E-mail:tangwenshi20@163.com, lixd@nju.edu.cn}
	\\
	$^1$School of Astronomy and Space Science, Nanjing University, Nanjing, 210023, People's Republic of China\\
	$^2$Key Laboratory of Modern Astronomy and Astrophysics, Nanjing University, Ministry of Education, Nanjing, 210023, People's Republic of China
}

\date{Accepted XXX. Received YYY; in original form ZZZ}
\pubyear{\today}

\begin{document}
\label{firstpage}
\maketitle

\begin{abstract}
The standard formation theory of binary millisecond pulsars (BMSPs) predicts efficient orbital circularization due to tidal interaction during the previous mass transfer phase. Therefore, BMSPs are expected to have a circular orbit. However, the discovery of several eccentric BMSPs (eBMSPs) with a white dwarf (WD) companion has challenged this picture. In particular, recent observation reveals that the spin angular momentum of the eBMSP J0955$-$6150 is tilted at an angle $>4.8^{\rm \degree}$ from the orbital angular momentum. This is the first time that a tilt angle is deduced for eBMSPs, which provides an important clue to their formation mechanism. Both the orbital eccentricity and tilt angle could be qualitatively accounted for by asymmetrical mass ejection during thermonuclear flashes from proto-WDs (so-called the thermonuclear rocket model), but detailed studies are still lacking. In this paper, we simulate the impact of the kick caused by asymmetrical mass ejection on the properties of BMSPs. We find that the thermonuclear rocket model can potentially explain the observational characteristics of both eBMSPs and normal BMSPs under reasonable input parameters. In addition, our results predict a wide range of the orbital period (from less than one day to more than several hundred days) for eBMSPs, which can be tested by future observations.
\end{abstract}


\section{Introduction}\label{Sect_Intro}
In the standard formation theory, binary millisecond pulsars (BMSPs) originate from low-mass X-ray binaries (LMXBs) which have undergone long time ($\sim 10^8-10^{10}$ yr) mass transfer \citep{1982Natur.300..728A,1982CSci...51.1096R, 1991PhR...203....1B}. In the LMXB phase, the neutron star (NS)'s spin is accelerated to ms periods and the binary orbit is circularized by strong tidal interactions. As a consequence, the resultant BMSPs are expected to reside in circular orbits \citep{1992RSPTA.341...39P}. In fact, the majority of BMSPs with Helium white dwarf (He WD) companions in the Galactic field have very small eccentricities in the range  of $10^{-7}-10^{-3}$, consistent with the theoretical predictions. However, recent observations reveal the existence of a population of eccentric BMSPs (eBMSPs), which challenges the standard formation theory. \cite{2008Sci...320.1309C} firstly reported the discovery of a BMSP J1903+0327 with a high eccentricity ($e\sim 0.44$) and a G-type main-sequence companion star. The current consensus is that this binary originated from the disruption of a triple system \citep{2011MNRAS.412.2763F, 2011ApJ...734...55P,2012MNRAS.424.2914P}.  
After that, several eBMSPs with He WD companions in the Galactic field were discovered \citep[see table 1 of ][for a list of eBMSPs]{2022A&A...665A..53S}. The properties of most eBMSPs can be briefly summarized as follows: (1) the eccentricities $e\sim 0.0274-0.14$; (2) similar orbital periods $\sim 20-30$\,days and WD companion masses $\sim 0.25-0.3\,M_{\rm \odot}$; (3) fully recycled pulsars with short spin periods $\sim 2-11$\,ms; and (4) a wide range of the neutron star (NS) masses $M_{\rm NS}\sim 1.3-1.8\,M_{\rm \odot}$. 

There are five models  proposed to explain (at least some of) eBMSPs in the literature, and they can be classified into three categories, i.e., the phase transition model \citep{2014MNRAS.438L..86F, 2015ApJ...807...41J}, the thermonuclear runaway burning model \citep{2014ApJ...797L..24A, 2021ApJ...909..161H}, and the resonant convection model \citep{2022MNRAS.509L...1G}.
In the phase transition model, the orbital eccentricity is caused by sudden mass loss due to the phase transition of the compact object after the LMXB phase - either accretion-induced collapse of a super-Chandrasekhar mass WD into an NS \citep{2014MNRAS.438L..86F}, or conversion of an NS into a strange quark star \citep{2015ApJ...807...41J}. 
In the thermonuclear runaway burning model, the eccentricity is induced by mass loss from the proto-WDs \citep{2014ApJ...797L..24A, 2021ApJ...909..161H}.  Because of unstable CNO shell burning on the surface of proto-WDs (after the LMXB phase and prior to the WD cooling stage), part of the ejected mass may form a circumbinary disk around the binary which dynamically interacts with the binary and pumps up an eccentricity \citep{2014ApJ...797L..24A}.   \cite{2021ApJ...909..161H} instead proposed that mass is asymmetrically lost from the proto-WDs, which imparts a kick on the binary and makes the binary orbit to be eccentric. 
In the resonant convection model, \cite{2022MNRAS.509L...1G} argued that a coherent resonance between the orbital period and the convective eddies in the red giant progenitor of the WD may drive the anomalous eccentricity in eBMSPs.

Currently, all the models are subject to various uncertainties and are not fully consistent with observations. For example, the phase transition models predict rather specific pulsar masses, which are disfavored by the wide range of the pulsar masses in eBMSPs. The thermonuclear runaway burning models struggle to explain why the orbital periods of eBMSPs are concentrated in a narrow range $\sim 20-30$\,days. For the resonant convection model, it is difficult to account for why some BMSPs with orbital periods similar to that of eBMSPs have not undergone resonant interactions and remained in circular orbits. 
Obviously, more observational clues are required to justify or put stringent constraints on the theories.

Recent observations of PSR J0955$-$6150 carried out by \cite{2022A&A...665A..53S} showed that the binary orbit is eccentric, and more interestingly, there is misalignment between the spin angular momentum (SAM) of the pulsar and the orbital angular momentum (OAM). The tilt angle  ($\mu$) between them is  $>4.8^{\degree}$ at 99\% CI. This is the first time that a tilt angle in BMSPs is reported. Detection of an SAM-OAM tilt angle is not surprising in other types of double degenerate systems, such as double NSs \citep[e.g., ][]{1989ApJ...347.1030W, 1998ApJ...509..856K}, double black holes (BHs), and BH$-$NS systems \citep{2021arXiv211103634T, 2021ApJ...915L...5A}. Misalignment in these systems is generally thought to be produced by the supernova events during the formation of the second compact object or the dynamical process in dense environment \citep[e.g. ][]{2000ApJ...541..319K, 2011ApJ...742...81F, 2016ApJ...832L...2R, 2017ApJ...846..170T,2017ApJ...846L..11L, 2018PhRvD..98h4036G}. In principle,  it is possible that an SAM-OAM misalignment exists in BMSPs in globular cluster via dynamical process. However, such misalignment is beyond the expectation of the standard formation theory for BMSPs in the Galactic field, because the old NSs should have experienced long duration of accretion \citep{1982Natur.300..728A, 1991PhR...203....1B}. Though the natal kick during the NS formation may initially excite an eccentricity and a tilt angle, the binary orbit must have been circularized by tidal interaction and the SAM and OAM must have been aligned by the accretion torques \citep{1983ApJ...267..322H, 1992RSPTA.341...39P, 2014MNRAS.439.2033G}. This means that the tilt angle in PSR J0955$-$6150 was very likely produced after the mass transfer had ceased. Unfortunately, most of the proposed formation channels of eBMSPs do not forecast such a large tilt angle \citep[see][for detailed discussion]{2022A&A...665A..53S}. In addition, PSR J0955$-$6150 has a He WD with mass $0.254(2)\,M_{\rm \odot}$, which is smaller than the theoretical prediction according to the relation between the WD mass and orbital the period \citep[the $M_{\rm WD}-P_{\rm orb}$ relation,][]{1999A&A...350..928T}. This is also difficult to understand in the standard theory.

 Interestingly, \cite{2022A&A...665A..53S} noticed that an SAM-OAM tilt angle may be qualitatively produced by the thermonuclear rocket model proposed by \cite{2021ApJ...909..161H}. This model assumes that the thermonuclear flash(es) on the surface of a proto-WD during its cooling sequence lead to asymmetrical mass ejection. As a consequence, the proto-WD suffers mild kicks, causing the host orbit to become eccentric and the SAM to be misaligned with the OAM. \cite{2022A&A...665A..53S} performed toy simulations of the thermonuclear rocket effect in single flash event with the kick velocity ($v_{\rm k}$) of 8\,km\,s$^{-1}$, but failed to obtain a large tilt angle and a long orbital period to match the observation of PSR J0955$-$6150. 
 
 However, numerical simulations \citep[e.g.][]{2001MNRAS.323..471A, 2016A&A...595A..35I} demonstrated that  there are likely multiple flashes on the surface of the proto-WDs depending on their masses.  After the Roche-lobe overflow (RLO) process, a thick hydrogen envelope (with mass $\sim 10^{-2}M_{\rm \odot}$) is left on the outer layer of the proto-WD \citep{2000MNRAS.316...84S, 2001MNRAS.323..471A, 2004MNRAS.352..249B, 2016A&A...595A..35I}. Subsequently, the proto-WD goes through a phase of contraction with almost constant luminosity and increasing effective temperature. In this phase, the luminosity is mainly supplied by the CNO burning in the shell. When the proto-WD starts cooling, the CNO burning is not sustained because the temperature in the burning shell is too low, and gravitational energy released due to contraction becomes the dominant energy source. As the temperature in the hydrogen shell rises to a critical value, the pp-chain burning is ignited, and then a flash starts with increasing luminosity. After the flash, the proto-WD contracts with an almost constant luminosity, and evolves following the above cycle. When the temperature of the hydrogen shell reaches the critical value again, a second flash occurs, and so on \citep[See][for a detailed description of the process of shell flashes]{2000MNRAS.316...84S}. In the simulation of \citet{2016A&A...595A..35I}, the maximum number of flashes can reach 26 if the effect of element diffusion (ED)\footnote{Element diffusion is the physical mechanism for mixing of chemical elements due to pressure, temperature and composition gradients \citep{1994ApJ...421..828T}.} is considered. Thus, it is crucial to test the thermonuclear rocket model by considering the influence of multiple flashes (or multiple kick events). 

The rest of this paper is organized as follows. In Sect.\,\ref{Sect_methods}, we introduce the assumptions and the simulation methods used in this work. We present our calculated results in Sect.\,\ref{Sect_results}, and summarize in Sect.\,\ref{Sect_sum}.
		
\section{Assumptions and Methods}\label{Sect_methods}
We consider binaries consisting of an NS and a proto-He WD, which is just detached from its RL and enters the cooling sequence. The initial value of $e$ and $\mu$ are both set to be 0. The initial orbital periods ($P_{\rm orb}^{\rm i}$) are generated from a uniform distribution. Then the WD masses are calculated according to the $M_{\rm WD}-P_{\rm orb}$ relation with a solar metallicity \citep[$Z=0.02$; ][]{1999A&A...350..928T}. Considering the fact that the measured masses of the NSs are in a broad range, we performed calculation with both $M_{\rm NS}=1.4$ and $2.0\,M_{\rm \odot}$, and found that there is only minor difference. Therefore, we only show the case of $M_{\rm NS}=1.4\,M_{\rm \odot}$ without loss of generality. 

The proto-WD is covered by a remaining hydrogen envelope. A significant part of hydrogen is burned stably at the bottom of the envelope, while the rest is consumed during shell flash(es) \citep[e.g.][]{2001MNRAS.323..471A, 2004MNRAS.352..249B, 2016A&A...595A..35I}. The released energy during flash(es) can heat the outer layer material and cause the inflation of the WD's envelope. \cite{2021ApJ...909..161H} argued that this energy may accelerate some outer layer material to a velocity of $5\times 10^{3}$\,km\,s$^{-1}$, comparable to the escape velocity of the WDs, and some of the envelope material is ejected asymmetrically from WDs. This idea originates from  \cite{2020MNRAS.492.3343S} who showed that asymmetrical mass loss during nova bursts is required to explain the decay of the orbit after a nova burst in some cataclysmic variables. Consequently, asymmetrical mass loss during shell flash(es) may impart a kick to the WD and result in an eccentric orbit. By using the law of momentum conservation, \cite{ 2021ApJ...909..161H} estimated the magnitude of the kick velocity to be around 8\,km\,s$^{-1}$, depending on the WD mass ($M_{\rm WD}$), the average ejected mass in each flash ($\Delta M_{\rm each}$), and the velocity of the ejecta ($V_{\rm ejct}$) (See Eq.\,[\ref{Eq_vk}]). In this work, we will examine the influence of kicks on the orbital period, the  eccentricity, and the  angle between the SAM and OAM simultaneously.

We calculate the eccentricity and the tilt angle after a flash using the method presented in \citet{2002MNRAS.329..897H}. Because the kick event may occur at arbitrary position in the orbit, in each simulation we randomly choose a mean anomaly $\mathcal{M}$ in $[0,2\pi]$ and solve the Kepler's equation for the eccentric anomaly $E$. The corresponding binary separation is $r=a(1-e{\rm cos}E)$, where $a$ is the semi-major axis. It is noted that, because our calculations involve multiple kicks and the orientation of the kick velocity is assumed to be randomly oriented in each flash, the orientations of OAM after several flashes are not always coplanar. Thus, the final tilt angle is not a simple addition or subtraction of the tilt angle after each kick event. In Appendix\,\ref{subsect_get_mu}, we present the method to calculate the tilt angle after multiple flashes.
 
An important issue in the thermonuclear rocket model is the efficiency of tidal circularization between and during flashes.
The proto-WD in the inter-flash phase is composed of a  degenerated He core and a non-degenerate, radiation-dominated hydrogen envelope with  mass $\sim 10^{-2}\,M_{\rm \odot}$. 
Using the {\tt MESA} code \citep{2011ApJS..192....3P, 2013ApJS..208....4P, 2015ApJS..220...15P,2018ApJS..234...34P}, we evolve a binary with $M_{\rm NS}^{\rm i}=1.4\,M_{\rm \odot},\, M_{\rm donor}^{\rm i}=1.4\,M_{\rm \odot}$, and $P_{\rm orb}^{\rm i}=$ 1.07\,days, until the formation of a BMSP with a 0.25\,$M_{\rm \odot}$ WD companion, roughly matching the observation of PSR J0955$-$6150. 
We obtain the  density and temperature profiles from the {\tt MESA}’s output files. Then we solve the equation of tidal torque presented in \citet[][Eq. 41]{1984MNRAS.207..433C} to evaluate the total tidal torque $T_{\rm tot}$ and the torque exerted by the degenerate He core ($T_{\rm He,\rm core}$). We find that $T_{\rm He, core}/T_{\rm tot}\sim 98\%$, i.e. the envelope only contributes a tiny fraction to the total torque. Consequently, we can safely use the tidal circularization timescale $\tau_{\rm circ}$ designed for double-degenerate binaries to estimate the efficiency of orbital circularization in the inter-flash phase \citep{1984MNRAS.207..433C, 2002MNRAS.329..897H}, in which the radiative damping mechanism is adopted for the WD's interior. We get $\tau_{\rm circ} \approx 1.5\times 10^{26}$\,yr for the above example and $4.0\times 10^{21}$\,yr for a binary consisting of a $\sim 0.20\,M_{\rm \odot}$ proto-WD. The extremely long circularization timescales indicate that tidal circularization is negligible in the inter-flash phase, which is consistent with the conclusion of \cite{2016A&A...595A..35I}.

We then consider tidal torque during the flashes. In this phase the proto-WD can dramatically inflate and possibly fill its RL with a duration up to $\sim 10^{\rm 2}-10^{3}$ yr \citep[e.g. ][]{2004MNRAS.352..249B}. 
The released energy during the flashes gives rise to a pulse-driven, deep convective zone within the hydrogen burning shell \citep{2016A&A...595A..35I}. Thus, equilibrium tide is more suitable in this situation (Yin Qin, private communication). Assume that the WD's spin synchronizes with the orbit, the evolution of $a$ and $e$ follows \citep{1981A&A....99..126H, 2002MNRAS.329..897H},
\begin{eqnarray}\label{eq_dedt}
\frac{de}{dt} & = & - 27 \left( \frac{k}{T} \right)
q \left( 1 + q \right) 
\left( \frac{R_{\rm WD}}{a} \right)^8 \frac{e}{\left( 1 - e^2 \right)^{13/2}} 
 \nonumber \\  & & \times 
\left[ f_3 \left( e^2 \right) - \frac{11}{18} \left( 1 - e^2 \right)^{3/2} 
f_4 \left( e^2 \right) \right],
\end{eqnarray}
and
\begin{eqnarray}\label{eq_dadt}
\frac{da}{dt} & = & - 6 \left( \frac{k}{T} \right) 
q \left( 1 + q \right) 
\left( \frac{R_{\rm WD}}{a} \right)^8 \frac{a}{\left( 1 - e^2 \right)^{15/2}} 
 \nonumber \\  & & \times 
\left[ f_1 \left( e^2 \right) - \left( 1 - e^2 \right)^{3/2} 
f_2 \left( e^2 \right) \right], 
\end{eqnarray}
where $f_{n}$s are the polynomial expressions in $e^{2}$ given by  \citet{1981A&A....99..126H}, $k$ the apsidal motion constant, and $T$ the timescale on which significant changes in the orbit take place. The mass ratio $q=M_{\rm WD}/M_{\rm NS}$, and the combined parameter of ($k/T$) can be expressed as \citep{1996ApJ...470.1187R}
\begin{equation}\label{Eq_k_d_T}
\left( \frac{k}{T} \right) = \frac{2}{21} \frac{f_{\rm conv}}
{{\tau}_{\rm conv}} \frac{M_{\rm Env}}{M} \, {\rm yr}^{-1}.
\end{equation}
Here the convection turnover timescale
\begin{equation}\label{Eq_tconv} 
{\tau}_{\rm conv} = 0.4311 \left[ \frac{M_{\rm Env} R_{\rm Env} \left( R_{\rm WD} - \frac{1}{2} 
R_{\rm Env} \right)}{3 L} \right]^{1/3} \, {\rm yr},
\end{equation}
where $M_{\rm Env}$ ($\sim 10^{-2}\,M_{\rm \odot}$) is the mass of the hydrogen-rich envelope, and $R_{\rm Env}$ the depth of the convective envelope. The luminosity $L$ of the proto-WD during the inflation phase evolves with time, and we take $L\sim 10^{\rm 2}\,L_{\rm \odot}$ according to the numerical calculations of \citet[][]{2004MNRAS.352..249B}. The numerical factor $f_{\rm conv}\le1$ and is simply set to be 1 here. Note that smaller $f_{\rm conv}$ implies longer tidal circularization time. In addition, we assume that the proto-WD fills its RL (with radius $R_{\rm L, WD}$) during the whole flash, and calculate the WD's radius with \cite{1983ApJ...268..368E}'s equation
\begin{equation}\label{Eq_rla}
\frac{R_{\rm WD}}{a} = \frac{R_{\rm L, WD}}{a} = \frac{0.49 q^{2/3}}{0.6 q^{2/3} + {\rm ln}\left( 1 + q^{1/3} \right)}.
\end{equation}
Because the radius of the degenerate He-core is much smaller and the inflated envelope dominates the size of the WD, we let $R_{\rm Env}=R_{\rm WD}=R_{\rm L, WD}$. We then incorporate the above equations into our code and let the tidal torque act on the binary for a time randomly between $10^{2}$ and $10^{3}$\,yr. Note that our treatment of $f_{\rm conv}$ and $R_{\rm WD}$ might cause over-estimate of the tidal efficiency.

\section{results}\label{Sect_results}

\subsection{Single kick model for eBMSPs}
\begin{figure*}
\centering
\includegraphics[width=0.7\textwidth]{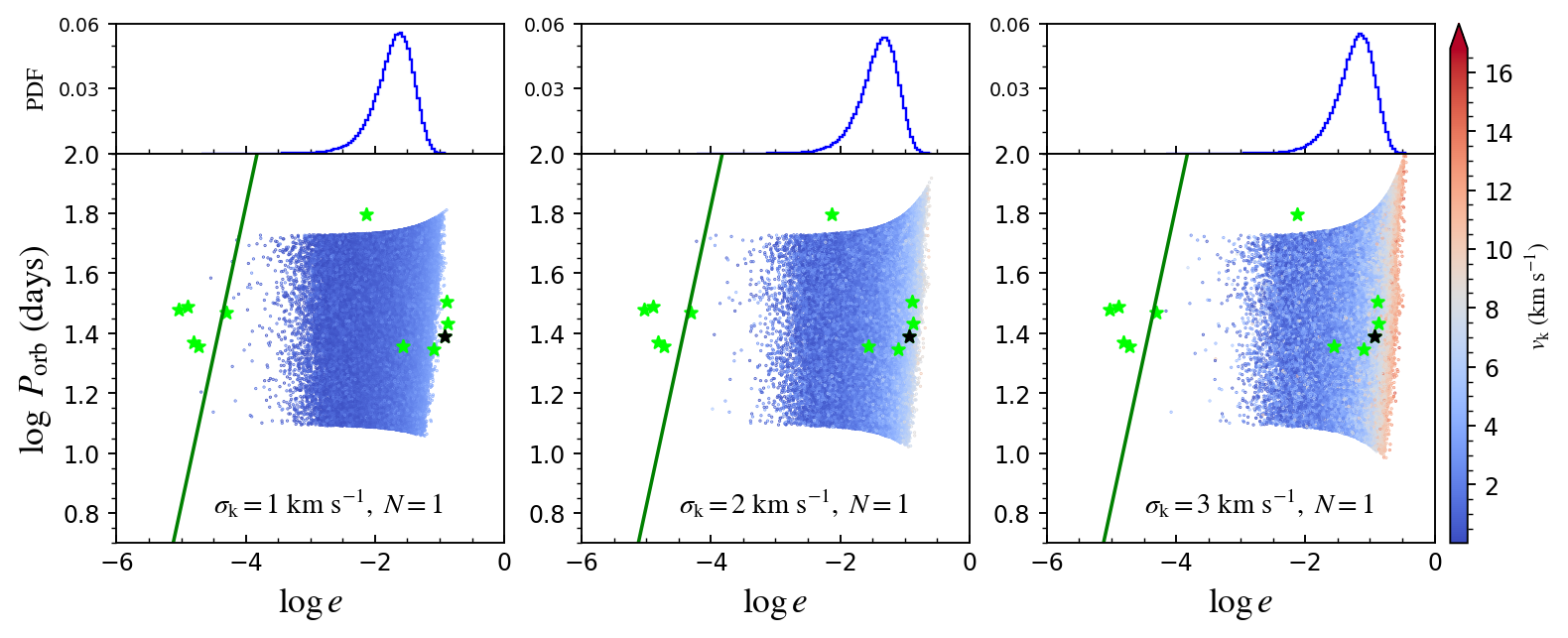}\\
\includegraphics[width=0.7\textwidth]{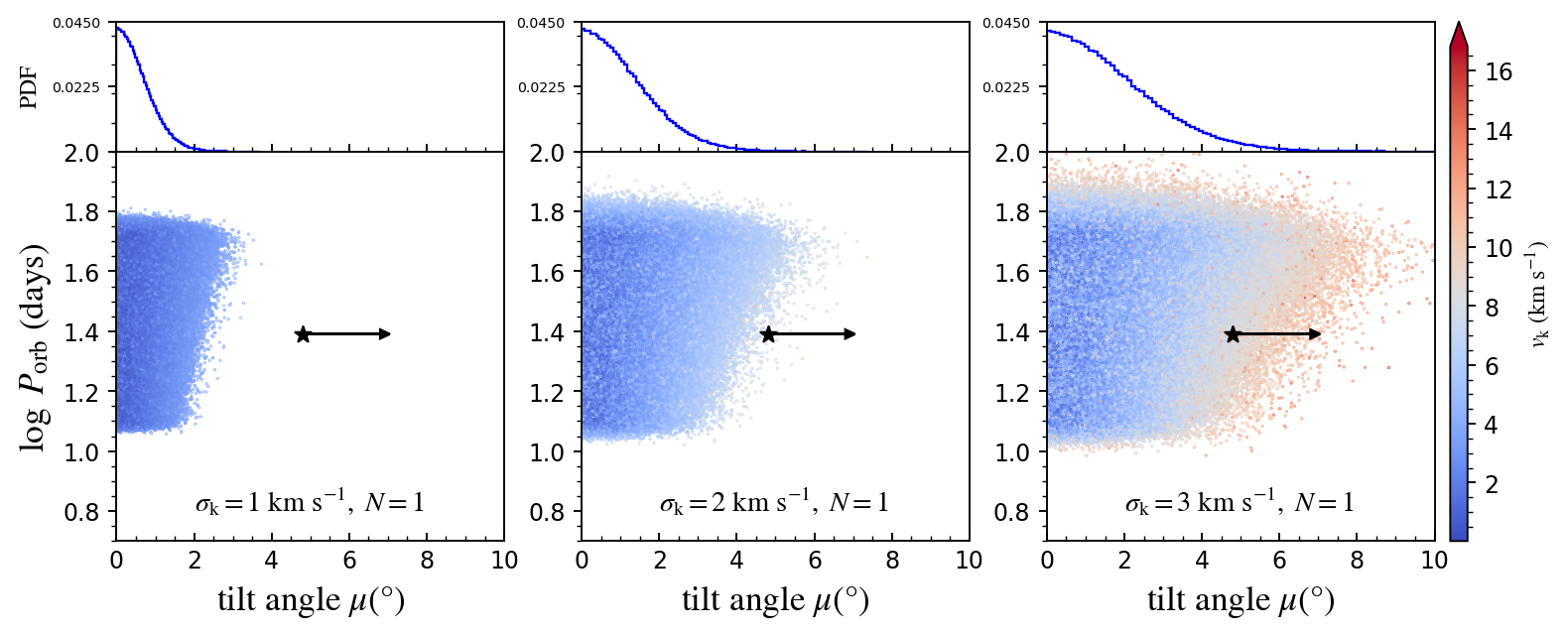}\\
\includegraphics[width=0.7\textwidth]{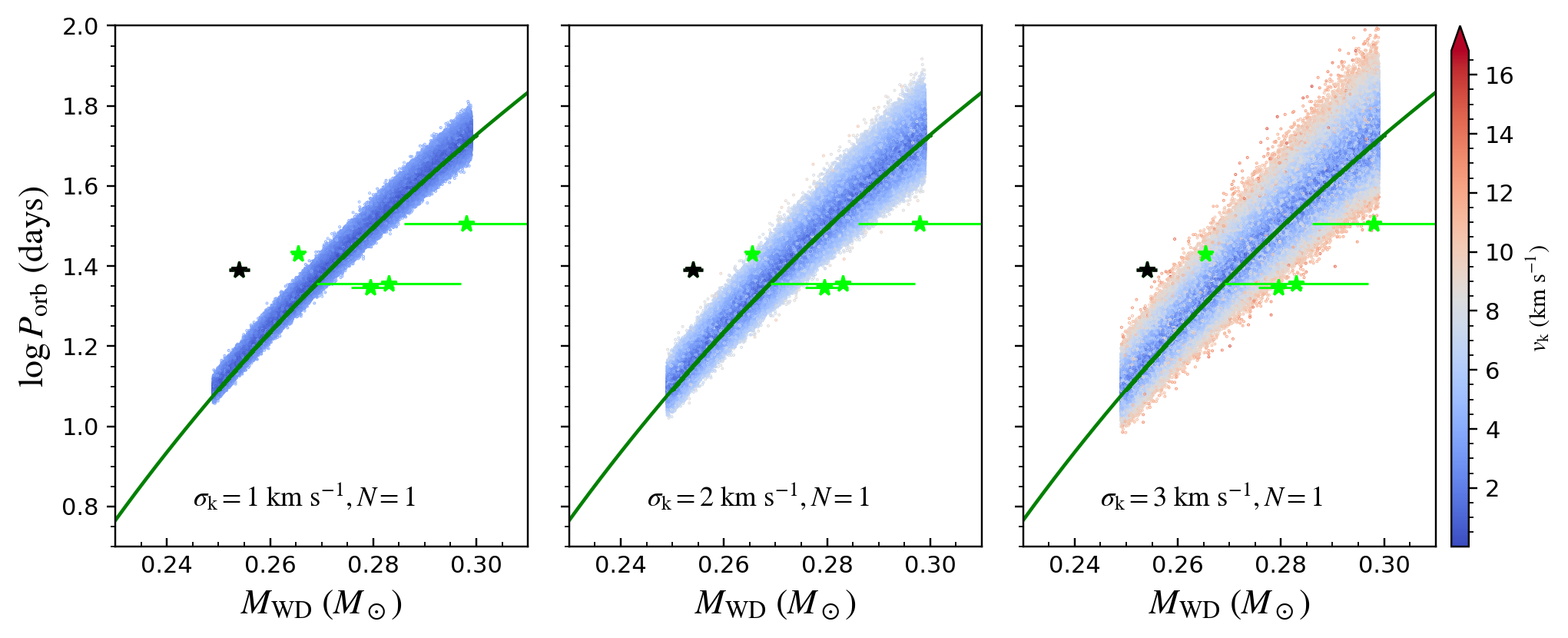}
 \caption{The parameter distribution of the simulated NS+He WD systems in the single kick model. The upper, middle, and bottom rows show the distribution of the orbital period as a function of the eccentricity, tilt angle and WD mass, respectively. In the upper and middle panels, we also show the probability distribution functions (PDFs) of $e$ and $\mu$, respectively. The left, middle, and right columns represent the cases of $\sigma_{\rm k}= 1$, 2 and 3 km\,s$^{-1}$, respectively. The green stars in the upper row mark the observed normal and eccentric BMSPs with orbital periods in the range of $\sim 20-65$ days. The black star marks PSR J0955$-$6150, and in the bottom row only eBMSPs are plotted. The green lines in the top and bottom rows are the theoretical $P_{\rm orb}-e$ and $P_{\rm orb}-M_{\rm WD}$ relations \protect\citep{1992RSPTA.341...39P,1999A&A...350..928T}, respectively. The color bars on the right of the panels represent the magnitude of $v_{\rm k}$.}
\label{Fig_Nkick_1}
\end{figure*}

We firstly present the predicted orbital properties of post-kick systems by considering a single kick event, which was also investigated by \cite{2022A&A...665A..53S}.  Their results with a constant $v_{\rm k}=8$\,km\,s$^{-1}$ produced too small tilt angles (always $<4.7^{\degree}$), and could not reproduce the observed orbital period of PSR J0955$-$6150. Here we adopt a slightly more sophisticated model assuming that the orientation of $v_{\rm k}$ is randomly distributed and the magnitude of $v_{\rm k}$ follows a Maxwellian distribution with standard deviation $\sigma_{\rm k}$. We consider three cases with $\sigma_{\rm k}=1$, 2, and 3\,km\,s$^{-1}$,  but with the same ejecta mass $\Delta M_{\rm ejct, tot} \,=10^{-3}M_{\rm \odot}$ \citep{ 2021ApJ...909..161H}, and simulate $5\times 10^{6}$ systems with the initial WD mass in the range of $[0.25,\,0.30]\,M_{\rm \odot}$ in each case. Fig.\,\ref{Fig_Nkick_1} illustrates the simulated distributions of $P_{\rm orb}$, $e$, and $\mu$. On the top of the $P_{\rm orb}-e$ and $P_{\rm orb}-\mu$ planes, we also show the probability distribution functions (PDFs) of $e$ and $\mu$, respectively.  The observed BMSPs are plotted with the green stars in the top panels, including both eBMSPs and normal BMSPs with orbital periods within $20-30$ days [that is, PSRs J0732+2314 \citep{2019ApJ...881..166M}, J1421$-$4409 \citep{2020MNRAS.496.4836S}, J1709+2313 \citep{2004ApJ...600..905L}, J1912$-$0952 \citep{2022arXiv220504232M}, and J2010+3051 \citep{2019ApJ...886..148P}].  The black star represents PSR J0955$-$6150. The green lines in the top and bottom panels  represent the theoretical $e-P_{\rm orb}$ \citep{1992RSPTA.341...39P} and $P_{\rm orb}-M_{\rm WD}$ \citep{1999A&A...350..928T} relations, respectively.

The top panels in Fig.\,\ref{Fig_Nkick_1} demonstrate that the distribution of the eccentricity shifts toward larger value with increasing $\sigma_{\rm k}$: the peak values are 0.016, 0.032, and 0.056 in the cases of $\sigma_{\rm k}=$1, 2, and 3\,$\rm km\,s^{-1}$, respectively. In all the three cases the simulated results can only partially cover the eBMSPs in the $P_{\rm orb}-e$ plane. In the middle panels, the tilt angle also exhibits the similar tendency.  Even when $\sigma_{\rm k}=3\,\rm km\,s^{-1}$, about $3\%$ systems can reach a tilt angle $\mu>4.8^{\rm \degree}$. 
Moreover, in the bottom panels, none of the simulated systems can match the orbital period of PSR J0955$-$6150. Therefore, we conclude that the single kick scenario seems not to be adequate for the formation of eBMSPs.

\begin{figure*}
\centering
\includegraphics[width=0.7\textwidth]{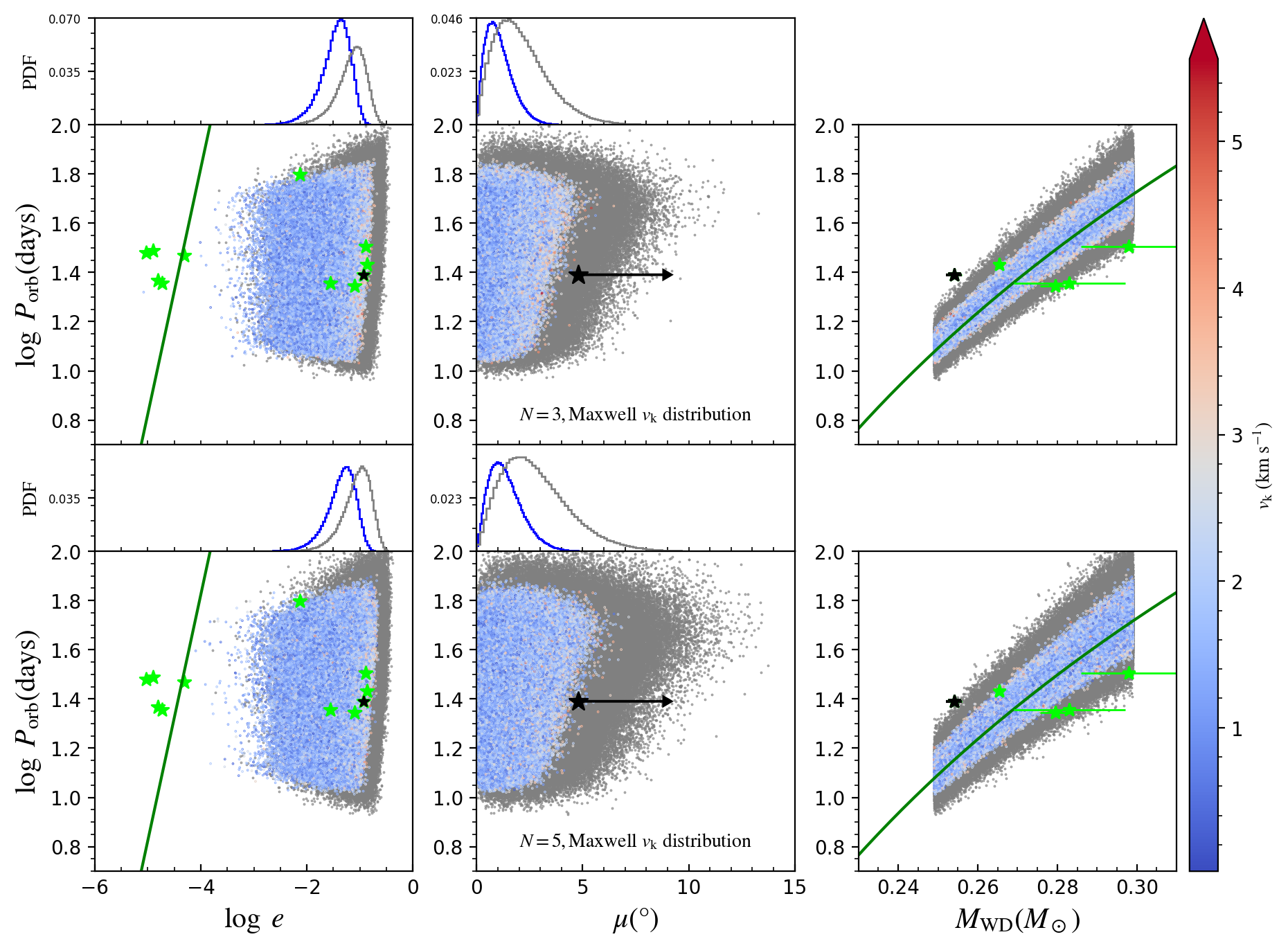}
\includegraphics[width=0.7\textwidth]{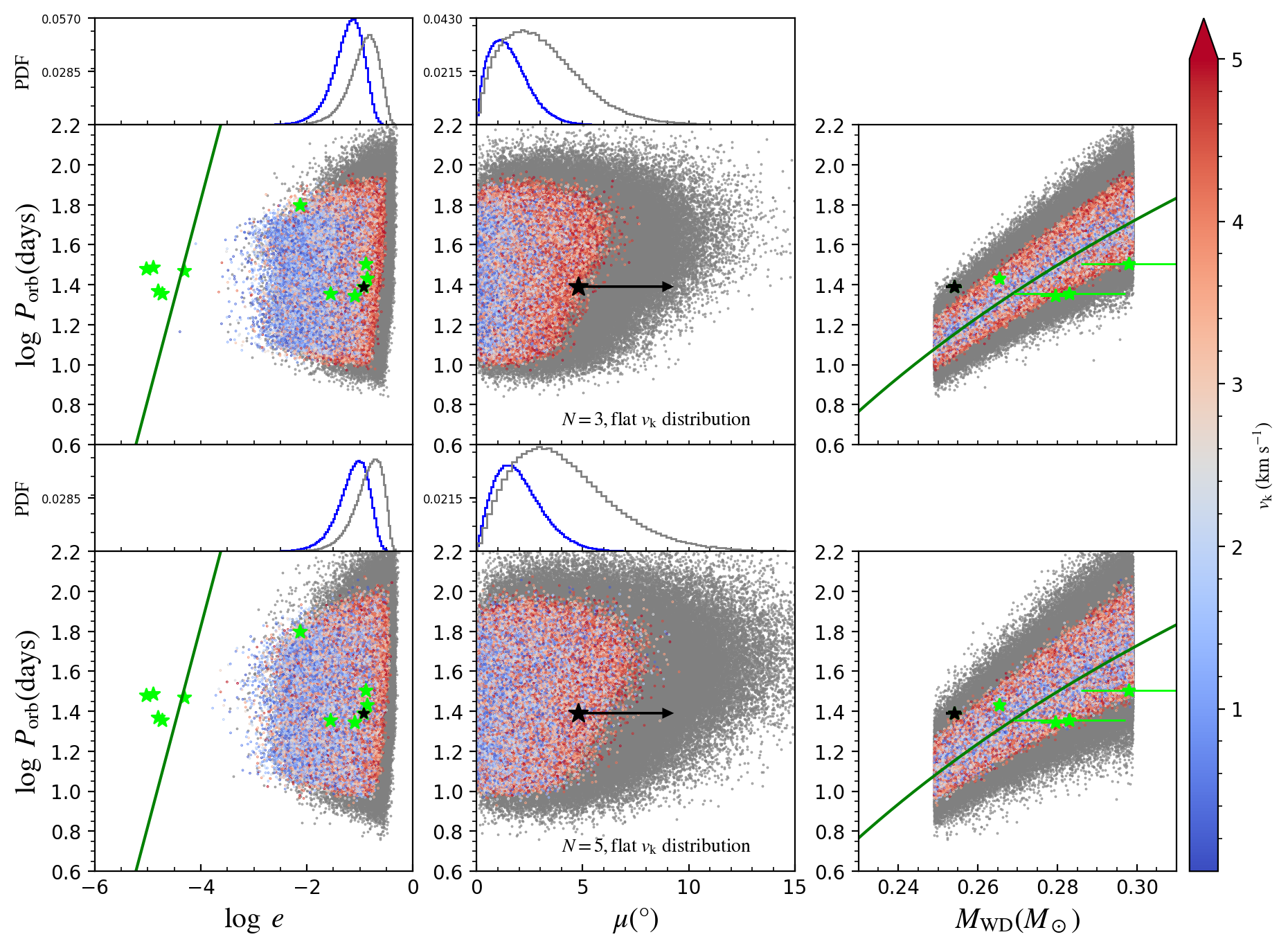}
 \caption{The parameter distribution of the simulated NS+He WD systems with multiple flashes. The magnitude and orientation of the kick velocity are both generated independently in each kick event. The kick velocity is assumed to follow Maxwellian distribution in the upper two rows and flat distribution in the lower two rows. For each distribution of $v_{\rm k}$, the upper and lower panels correspond to $N=3$ and 5, respectively. The values of $v_{\rm k}$ are limited to 5\,km\,s$^{-1}$ (colorful dots) or 10\,km\,s$^{-1}$ (grey dots). The PDFs are shown on the top of each panel.}
\label{Fig_Nkick_35_random}
\end{figure*}

\subsection{Multiple kick model for eBMSPs }\label{Result_Mult_Kick}

As pointed out before, a proto-WD likely undergoes multiple flashes depending on its mass. Previous numerical simulations show that the maximum number of flashes can reach 26 if the effect of ED is considered, although in most proto-WDs the number ($N$) of flashes is no more than $5-6$ \citep{2016A&A...595A..35I}. Thus, in this subsection we investigate the possible influence of multiple kicks on the orbital evolution. In the following calculations, we set $N=3$ and 5, and the total ejected mass $\Delta M_{\rm ejct, tot}= 10^{-3}\,M_{\rm \odot}$. We assume that in each flash, the ejected mass $\Delta M_{\rm each}$ = $\Delta M_{\rm ejct, tot}/N$ and the kick velocity is randomly orientated.

An example of our simulated results is shown in Fig.\,\ref{Fig_Nkick_35_random}. Here the magnitude of the kick velocity is assumed to follow a Maxwellian distribution (with $\sigma_{\rm k}=1$ and 2 km\,s$^{-1}$) or flat distribution (with $v_{\rm k}$ lying between 0 and 10 km\,s$^{-1}$). The upper two panels correspond to the Maxwellian distribution of $v_{\rm k}$ with $N=3$ and 5, and the lower two panels the flat distribution of $v_{\rm k}$ with $N=3$ and 5, respectively. The PDFs are also demonstrated on the top of each panel. The colorful and grey dots (and the blue and grey curves) stand for the post-kick systems with the last kick velocity $\leq 5$ and 10 km\,s$^{-1}$, respectively. 

A comparison of Figs.\,\ref{Fig_Nkick_1} and \ref{Fig_Nkick_35_random} shows that multiple kicks indeed produce larger eccentricities and tilt angles. For example, for Maxwellian distribution of $v_{\rm k}$ with $\sigma_{\rm k}= 2 \,\rm km\,s^{-1}$ and $N=5$, 69.7\% of the systems have an eccentricity within the observed interval ($0.0274-0.1345$) of eBMSPs, and 11.85\% of the systems have a tilt angle $\mu>4.8^{\degree}$. 
We also see that flat $v_{\rm k}$ distribution can produce slightly more systems with large eccentricities and tilt angle than Maxwellian distribution,
because flat distribution tends to generate more systems with relatively high $v_{\rm k}$. 
In addition, the increase in $N$ leads to larger change in the orbital period. By comparing with the observations, we conclude that multiple flashes with both Maxwellian and flat distributions of the kick velocity seem possible to reproduce the properties of eBMSPs.

We then construct a number of models by changing the key parameters that describe the kick properties, including the number of flashes ($N$) and the kick velocity distribution $\sigma(k)$ in various flashes. The meanings of the model names are summarized in Table 1. For example, model $N5\_\sigma(-2)\_A$ means that there are 5 flashes, and in each flash, the kick velocity independently follows flat distribution between 0 and 10 km\,s$^{-1}$, and model $N3\_\sigma(+1)\_B$ means that there are 3 flashes, and in all flashes the kick velocity follows the same Maxwellian distribution with the standard deviation $\sigma_{\rm k}=1$ km\,s$^{-1}$.

\begin{figure*}
\centering
\includegraphics[width=0.85\textwidth]{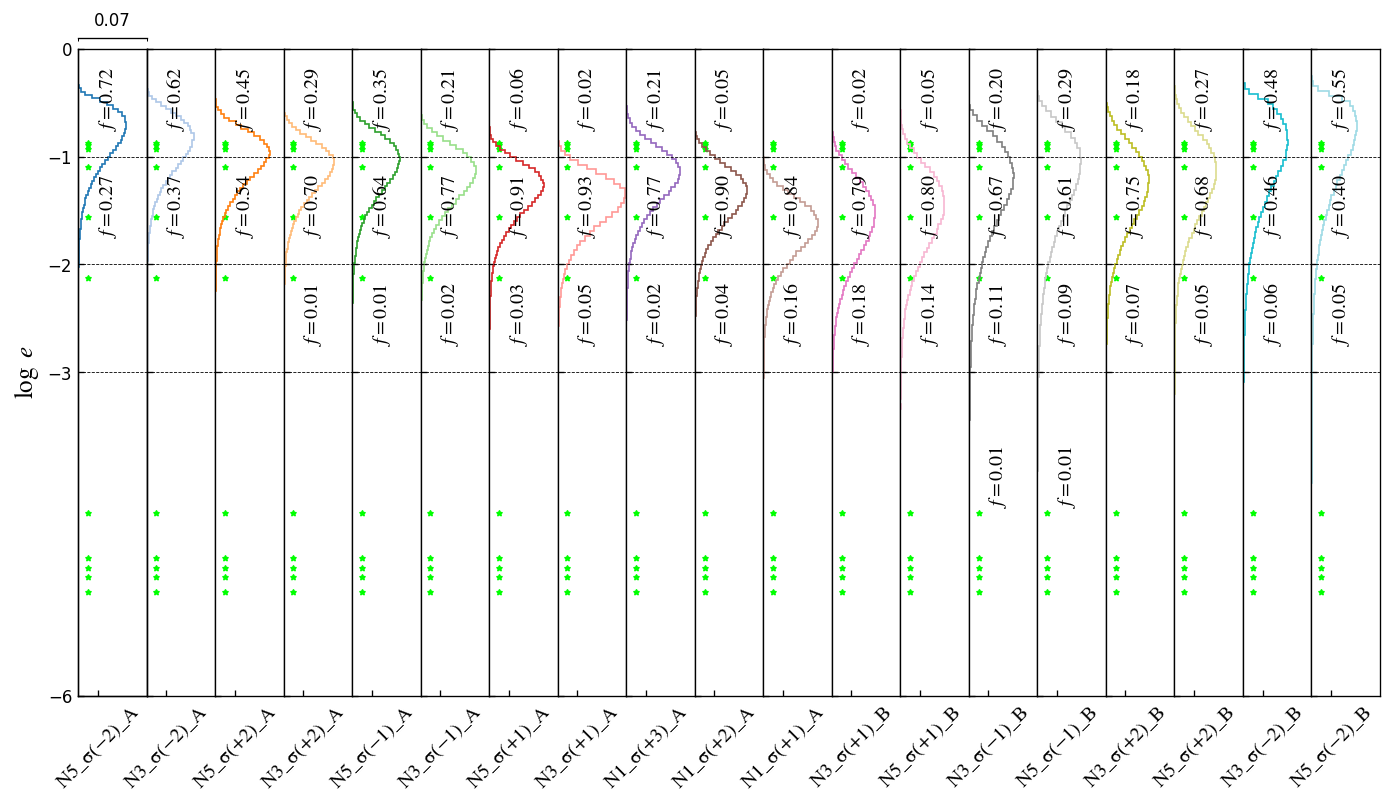}\\
\vspace{0.1cm}
\includegraphics[width=0.85\textwidth]{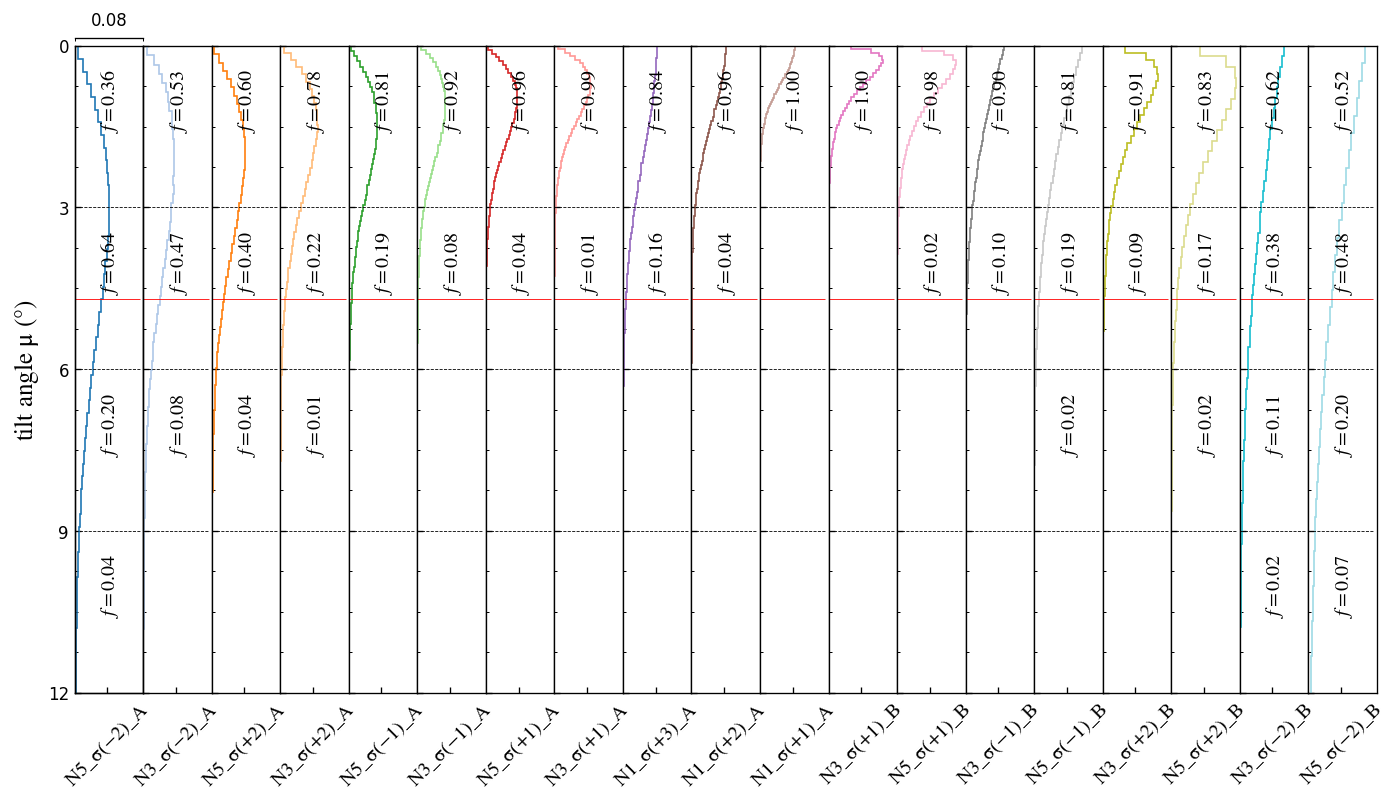}\\
 \caption{The PDFs of $e$ (upper panel) and $\mu$ (lower panel) for 19 representative models. In the upper panel, the green stars represent the observed eccentricities of BMSPs. In the lower panel, the red line represents the lower limit of the tilt angle of PSR J0955$-$6150. The values of $f$ denote the percentages of the systems located in specific intervals.}
\label{Fig_e_mu_distri_nofa}
\end{figure*}

In Fig.\,\ref{Fig_e_mu_distri_nofa}, we compare the simulated PDFs of the eccentricity (top panel) and the tilt angle (bottom panel) for 19 representative models. We also list the percentage ($f$) of systems with $e$ or $\mu$ in specific intervals confined by the dashed lines. The green dots in the top panel represent eBMSPs and normal BMSPs, and the red line in the bottom panel denotes the lower limit of the tilt angle of PSR J0955$-$6150. More detailed information about the models and the corresponding results are presented in Table\,\ref{table_results}.

From Fig.\,\ref{Fig_e_mu_distri_nofa} and Table\,\ref{table_results} we see that most of the multiple kick models can more or less cover the observed $e$ distribution of eBMSPs. However, to reach a tilt angle $\mu>4.8\degr$ with a sufficiently high probability requires relatively large $v_{\rm k}$ (with $\sigma_{\rm k}\ge 2$ kms$^{-1}$ or $v_{\rm k}\in[0,10]$ kms$^{-1}$) and/or $N$ ($\ge 3$).

\begin{table*}
	\small
	\begin{center}
	 	\caption{Description of the model parameters}
		 \label{table_model}
  		\begin{tabular}{cc}
		\hline 
				 N$i$            &  \makecell[l]{$i$ denotes the number of flashes.}\\
				 	\hline
		$\sigma(k)$         &    \makecell[l]{if $k>0$, $v_{\rm k}$ follows Maxwellian distribution with the standard deviation $\sigma_{\rm k}=k$\,$\rm km\,s^{-1}$; \\
				        if $k<0$, $v_{\rm k}$ follows flat distribution, with $k =-1$ and $-2$ corresponding to $v_{\rm k}\in(0,5)$ and $(0,10)$\,$\rm km\,s^{-1}$, repsectively. }\\
				\hline
	    Case &   \makecell[l]{A: the $v_{\rm k}$ distribution is independently set in each flash.\\
	    B: the $v_{\rm k}$ distributions in all flashes take the same form.}\\
		\hline
		\end{tabular}
	\end{center}
\end{table*}

\begin{figure*}
\centering
\includegraphics[width=0.85\textwidth]{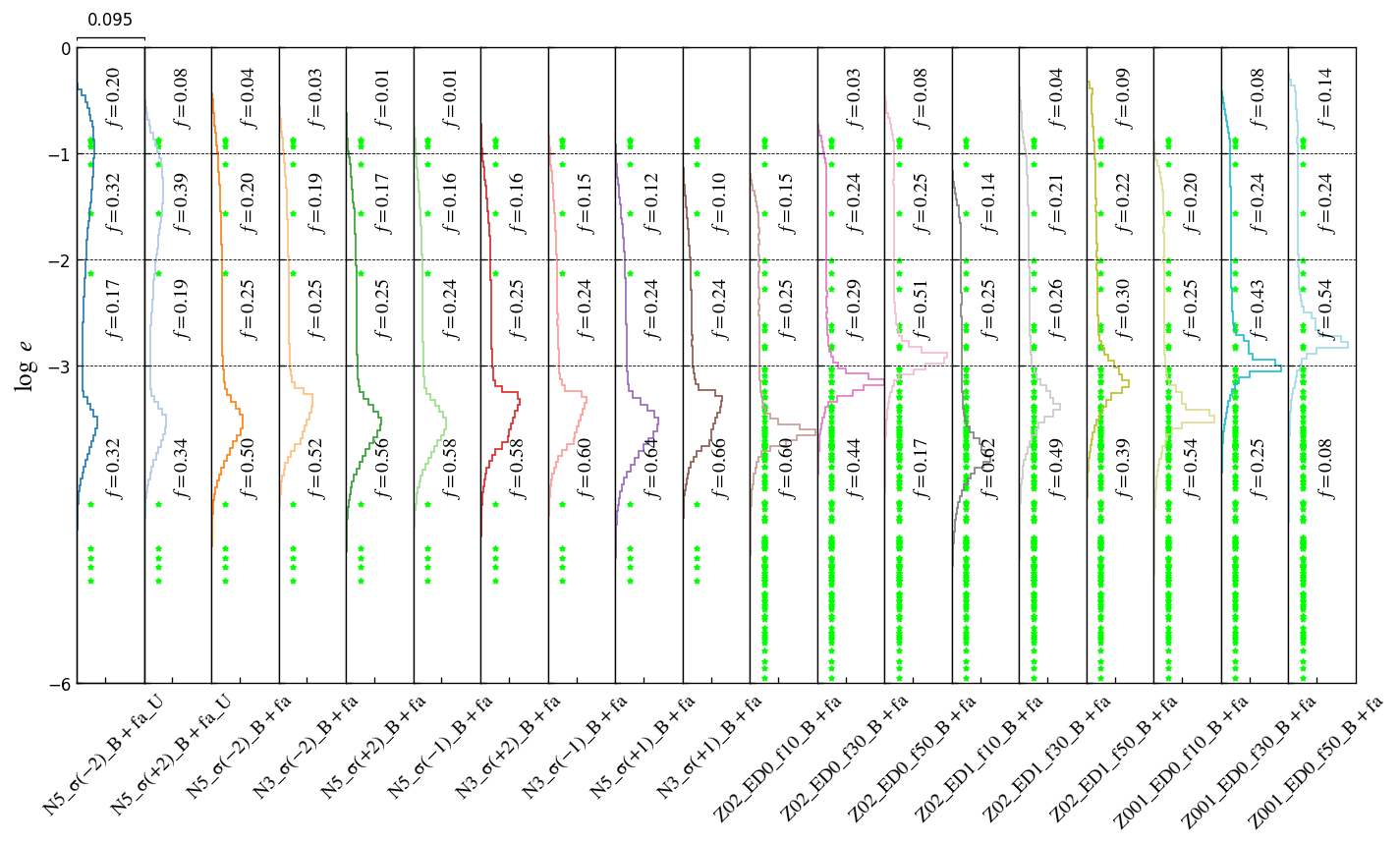}\\
\vspace{0.1cm}
\includegraphics[width=0.85\textwidth]{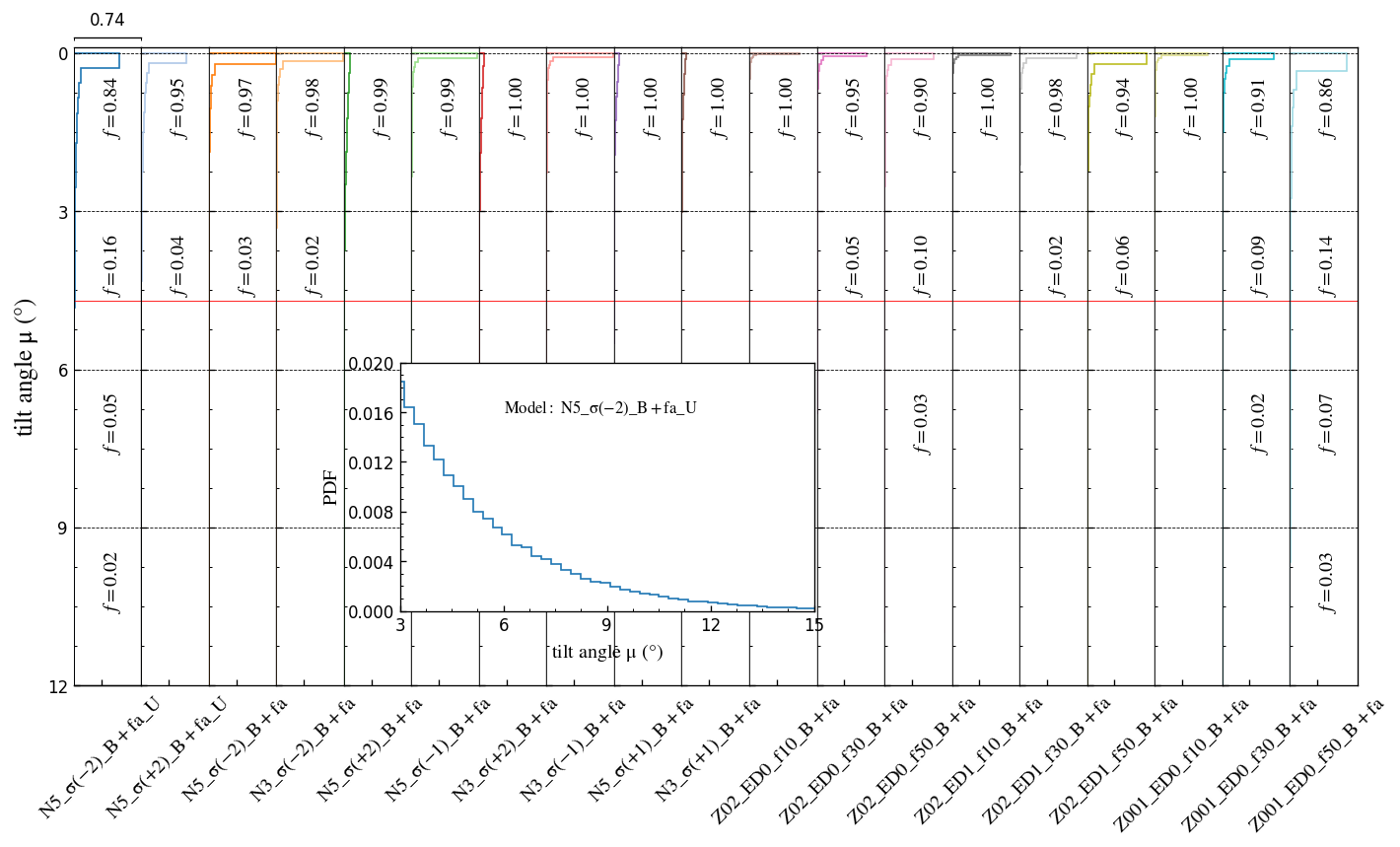}\\
 \caption{Similar to Fig.\,\ref{Fig_e_mu_distri_nofa}, but with the asymmetry parameter $f_{\rm a}$ considered in the left 10 columns, and with fitted relations considered in the right 9 columns. The inset in the lower panel shows the PDF in model $N5\_\sigma (-2)\_B+fa\_U$.}
\label{Fig_e_mu_distri_withfa}
\end{figure*}

\subsection{Multiple kick model for both BMSPs and eBMSPs }

A successful kick model should be able to simultaneously account for both eBMSPs and normal BMSPs. While the above models seem to be able to explain the properties of eBMSPs, none of them can cover the eccentricity distribution of normal BMSPs. 
This means that the realistic situations of shell flashes are more complicated than our assumptions about the kick velocity, that is, the kick velocity may depend on not only the ejecta mass but also the asymmetry of mass ejection. So we introduce an asymmetry parameter $f_{\rm a}$ which quantifies the extent of asymmetry in the mass loss processes during shell flashes. The case of $f_{\rm a}=1$ corresponds to fully asymmetric ejection, while $f_{\rm a}=0$ represents fully symmetric mass ejection without any kick. Then the modified kick velocity is  $f_{\rm a} v_{\rm k}$. Since the form of $f_{\rm a}$ is unclear, 
we tentatively consider two possibilities: (1) $\log\,f_{\rm a}$ follows a flat distribution in the range of $[-5,0]$, and (2) $\log f_{\rm a} = C(1-x)^{3}$, where $C=-5$ and $x$ is a random number in $[0,1]$. While in the former case the $f_{\rm a}$ distribution is dominated by relatively small values, in the latter case it is clustered around both small and large values. We add ``fa" and ``fa\_U" to the original model name to identify them, respectively.

We show the calculated results for 10 representative models with asymmetrical mass ejection considered in the left 10 columns of Fig.\,\ref{Fig_e_mu_distri_withfa} and in Table B1. Comparing with Fig.\,\ref{Fig_e_mu_distri_nofa} we find that the distributions of the eccentricity cover a wider range than predicted by models without $f_{\rm a}$ considered, but the tilt angle is clustered smaller value. Among all the models, model $N5\_\sigma (-2)\_B+fa\_U$, seems to simultaneously match the eccentricities of eBMSPs and normal BMSPs and the tilt angle of PSR J0955$-$6150. The inset in the lower panel shows the PDF of $\mu$ in detail.

\subsection{The effect of flash number and ejected mass}
We finally discuss the possible dependence of the flash number and ejected mass on the WD mass, rather taking them as fixed values. Depending on metallicity and ED, different WDs may experience different times of flashes \citep{2001MNRAS.323..471A,2016A&A...595A..35I}. Generally $N$ is smaller for more massive WDs with the same metallicity \citep{2016A&A...595A..35I}. In addition, the ejected mass during flashes may be different for different WDs. For example, \citet{2004MNRAS.352..249B} found that only $\sim$10\% of the hydrogen envelope mass ($M_{\rm Env, det}$) is lost for a $0.21\,M_{\rm \odot}$ WD during the flashes, while $\sim43\%$ of $M_{\rm Env, det}$ is lost for a $0.3\,M_{\rm \odot}$ WD in \citet{1986ApJ...311..742I}. 

Basing on the calculated data provided in Table A1 of \cite{2016A&A...595A..35I}, we derive an empirical $N-M_{\rm WD}$ relation by utilizing polynomial fitting (See Appendix\,\ref{subsec_fitN} for details). Then the average lost mass $\Delta M_{\rm each}$ during each flash is taken to be 
\begin{equation}\label{Eq_M_each}
\Delta M_{\rm each}=\frac{\Delta M_{\rm ejct, tot}}{N(M_{\rm WD})},
\end{equation}
where $\Delta M_{\rm ejct, tot}$ is also fitted as a function of $M_{\rm WD}$ by using the data of $M_{\rm Env, det}$ in \cite{2016A&A...595A..35I}, i.e.,
\begin{equation}
\Delta M_{\rm ejct, tot}(M_{\rm WD})=f_{\rm ejct}M_{\rm Env, det}(M_{\rm WD}).
\label{Eq_ejct_tot}
\end{equation}
Here $f_{\rm ejct}$ denotes the percentage of mass loss to $M_{\rm Env, det}$ during flashes, and is set to be randomly in the range of $[10\%,50\%]$ according to  \cite{1986ApJ...311..742I} and  \cite{2004MNRAS.352..249B}.
Then the kick velocity is calculated with \citep[see also][]{2021ApJ...909..161H}
\begin{equation}\label{Eq_vk}
v_{\rm k} = 8\,{\rm km\,s^{-1}}f_{\rm a}\left(\frac{\Delta M_{\rm each}}{10^{-3} M_{\sun}}\right)\left(\frac{V_{\rm ejct}}{\rm 2.5\times 10^3\,km\,s^{-1}}\right)\left(\frac{M_{\rm WD}}{0.3 M_{\sun}}\right)^{-1},
\end{equation}
where the ejecta velocity $V_{\rm ejct}$ is set to be $2.5\times 10^{3}$\,km\,s$^{-1}$. We assume that the magnitude of $v_{\rm k}$ is the same for all flashes but its orientation is randomly set during each flash. We consider three cases: (1) the metallicity $Z=0.02$ without ED (named as Z02\_ED0), (2) $Z=0.02$ with ED (Z02\_ED1), and (3) $Z=0.001$ without ED (Z001\_ED0). A log-normal distribution is adopted for $f_{\rm a}$. 
The simulated results are presented in the right 9 columns of Fig.\,\ref{Fig_e_mu_distri_withfa} and Table B2. In the model names, $f10$, $f30$, and $f50$ correspond to $f_{\rm ejct}=10\%$, $30\%$, and $50\%$, respectively. Note that here we include all systems with the occurrence of shell flash(es). Therefore, the range of $M_{\rm WD}$ is significantly wider than the  adopted range of $[0.25,0.30]\,M_{\sun}$ in previous discussion. Specifically, $M_{\rm WD}$ is in $[0.213,0.305]\,M_{\sun}$, $[0.167,0.393]\,M_{\sun}$ and $[0.252,0.350]\,M_{\sun}$ for models of Z02\_ED0, Z02\_ED1, Z001\_ED0, respectively. Thus, we plot all BMSPs with known eccentricities for comparison.  

The right 9 columns in Fig.\,\ref{Fig_e_mu_distri_withfa} show the following features. 

(1) When $f_{\rm ejct}=10\%$, the kick velocities can barely explain the observed eccentricities of eBMSPs. The predicted tilt angles ($<3^{\degree}$) are also smaller than the lower limit of $\mu$ for PSR J0955$-$6150. Increasing $f_{\rm ejct}$ can produce more systems with eccentric orbits and larger tilt angles. We find that models Z02\_ED0\_f50\_B+fa, Z001\_ED0\_f30\_B+fa, and Z001\_ED0\_f50\_B+fa seem to better match the observational properties of both eBMSPs and normal BMSPs.

(2) When ED is considered, both the eccentricities and tilt angles shift to smaller values compared with those without ED, due to relatively smaller ejected mass and kick velocity.

(3) Models with lower metallicity generally produce larger eccentricities and tilt angles under the same $f_{\rm ejct}$. This difference originates from the more massive envelope (i.e., larger $M_{\rm Env, det}$) for lower metallicity stars \citep{2016A&A...595A..35I}, which in turn results in larger kick velocity (See  Fig.\,\ref{Fig_vk_PDF_fit_DM_f50}).


\section{Summary}\label{Sect_sum}
In this work, we have investigated the feasibility of the thermonuclear rocket model for the formation of eBMSPs. Here, we focus on how to explain the tilt angel between the SAM and the OAM and the orbital period (or the WD mass) deviation from the theoretical prediction for PSR J0955$-$6150. We also explore the possibility to simultaneously explain the overall eccentricity distribution of both eBMSPs and  normal BMSPs under the thermonuclear rocket model. 

We first let all WDs have the same number $N$ of flashes and the same ejecta mass $\Delta M_{\rm ejec,tot}$ during flashes, and artificially limit the mass range of WDs within that of observed eBMSPs. Under these simplified assumptions, we show that the thermonuclear rocket model with multiple kicks could better reproduce the properties of both eBMSPs than with single kick. However, to account for the normal BMSPs simultaneously, one needs to carefully take into consideration the effect of asymmetry of mass ejection. 

We then discuss the more realistic case based on the numerical data of shell flashes on proto-WDs in previous studies. In this case, both $N$ and $\Delta M_{\rm ejec,tot}$ depend on $M_{\rm WD}$, metallicity and ED. We fit the relations among these factors and incorporate them into our simulations. We find that under such situation, a few models with relatively large (30\% - 50\%) mass ejection fraction can potentially match the observations. Moreover, our results predict a wide range of the orbital period (from less than one day to more than several hundred days) for eBMSPs, and some of them may have a tilt angle up to $\sim 10\degr$. We expect that these predictions to be tested by future observations.


\section*{Acknowledgments}
\quad We thank the referee, Pablo Marchant, for constructive comments and suggestions that have helped improve the manuscript. This work was supported by the National Key Research and Development Program of China (2021YFA0718500), the Natural Science Foundation of China under grant No. 12041301, 12121003, and Project U1838201 supported by NSFC and CAS.

\section*{Data Availability}
\quad All data underlying this article will be shared on reasonable request to the corresponding authors.

\bibliographystyle{mnras}

\appendix
\section{}\label{Sect_apdix}

\subsection{The tilt angle after multiple kicks}\label{subsect_get_mu}
\begin{figure}
\centering
\includegraphics[width=0.5\textwidth]{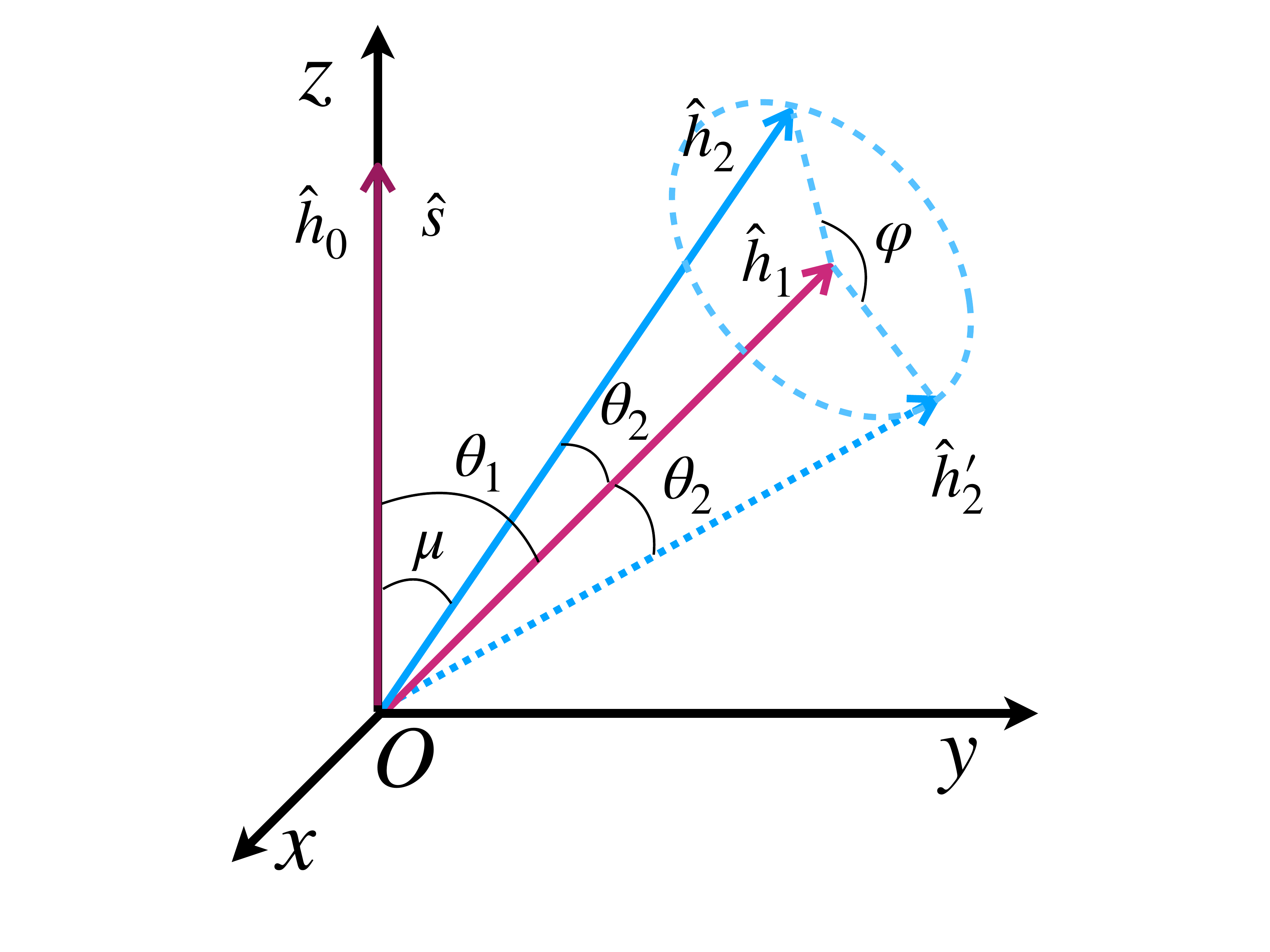}
 \caption{The geometry of the model for calculating $\mu$ after multiple kicks.}
\label{Fig_Rotat_vector}
\end{figure}

The geometry of the binary is shown in Fig.\,\ref{Fig_Rotat_vector}. The NS is at the coordinate origin, and the orbit is on the $xy$ plane.
We assume that the unit vector $\hat s$ of the spin momentum of the NS is initially aligned with the orbital angular momentum $\hat h_0$ and the $z$-axis, i.e., $\hat s=\hat h_0=(0,0,1)$.

After the first kick, the angle $\mu$ between $\hat s$ and $\hat h_1$ is $\mu_1=\theta_1=\langle\hat s,\hat h_1\rangle$. The toroidal angle of $\hat h_1$ is arbitrarily set, and we assume that ${\hat h}_1$ is in the $yOz$ plane.

After the second kick, the orbital angular momentum vector becomes $\hat h_2$. Using the method in Appendix\,A1 of \citet{2002MNRAS.329..897H}, we can obtain the angle between $\hat h_{1}$ and $\hat h_2$ , i.e., $\theta_2$. Since the angular momentum vector $\hat h_2$ is not coplanar with $\hat h_0$ and $\hat h_1$, the final tilt angle is not linear addition or subtraction between $\theta_1$ and $\theta_2$. We derive $\mu$ after the second kick in the following way. We first rotate $\hat h_1$  by the angle $\theta_2$ to get $\hat h'_2$ in the $yOz$ plane. Then, we use the Rodrigues’ Rotation Formula
\begin{equation}
\hat h_2=\cos\varphi\hat h'_2+(1-\cos \varphi)(\hat h'_2\cdot\hat h_1)\hat h_1+\hat h_1\times\hat h'_2\sin\varphi,
\end{equation}
to rotate $\hat h'_2$ along $\hat h_1$ by an arbitrary angle from a uniform distribution of $[0, 2\pi]$ and finally get $\hat h_2$. So, the angle between the current orbital angular momentum and the spin angular momentum of the NS is $\mu=\langle\hat s,\hat h_2\rangle$ (which is $\leq\theta_1+\theta_2$). For multiple kicks, $\mu$ can be obtained by repeating the above steps.

\subsection{Fitting the $N-M_{\rm WD}$ and $\Delta M_{\rm ejct,tot}-M_{\rm WD}$ relations }\label{subsec_fitN}
\begin{figure}
\centering
\includegraphics[width=0.5\textwidth]{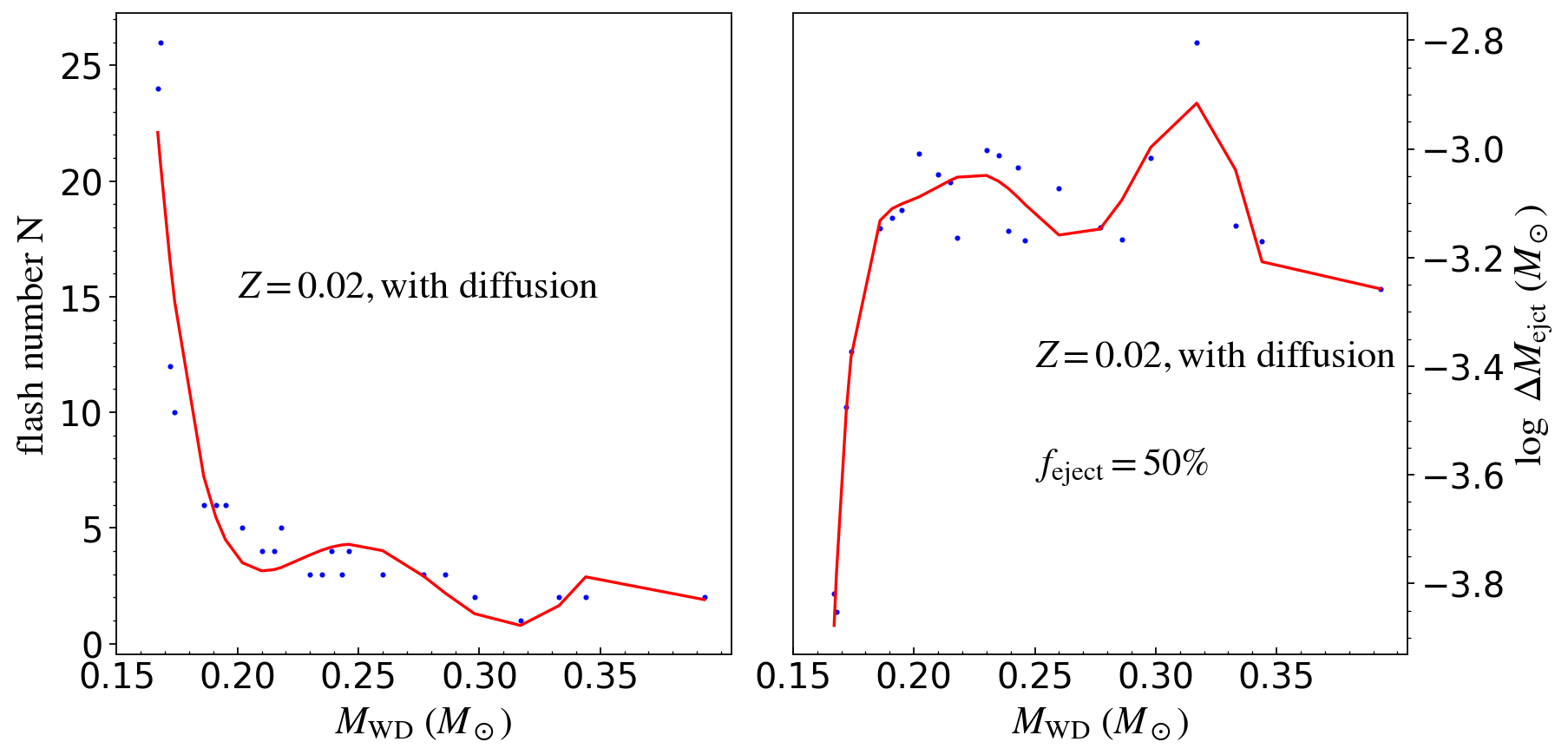}
 \caption{An example of the empirical $N-M_{\rm WD}$ (left) and $\Delta M_{\rm ejct,tot}-M_{\rm WD}$ (right) relations for model of Z02\_ED1. }
\label{Fig_N}
\end{figure}

Based on the data provided in Table A1 of \cite{2016A&A...595A..35I}, we utilize the polynomial function
\begin{equation}
F(x) = \sum_{\rm j=0}^{\rm n} w_{\rm j}x^{\rm j}
\end{equation}
to fit the $N-M_{\rm WD}$ and $\Delta M_{\rm ejct,tot}-M_{\rm WD}$ relations with the {\tt polyfit} function in the {\tt python} code, where $n=5$ for $N(M_{\rm WD})$ and 8 for $M_{\rm ejct, tot}(M_{\rm WD})$. An example is demonstrated in Fig.\,\ref{Fig_N}.

\begin{figure}
\centering
\includegraphics[width=0.5\textwidth]{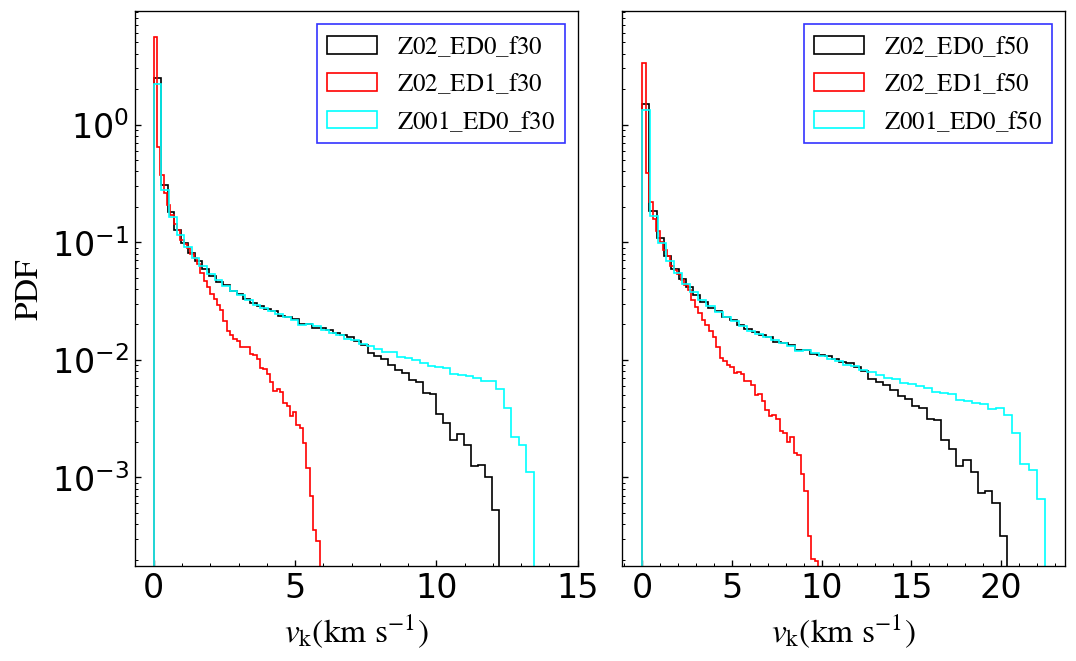}
 \caption{The obtained $v_{\rm k}$ distribution by using the fitting result of Appendix.\,\ref{subsec_fitN} and Eq.(\,\ref{Eq_vk}) to calculate $v_{\rm k}$, where a log-normal $f_{\rm a}$ distribution has been taken into account. The left and right panels show the case with $f_{\rm ejct}=$ 30\% and 50\%, respectively. In each panel, the green, black and red lines correspond to the models of Z001\_ED0, Z02\_ED0, Z02\_ED1, respectively.}
\label{Fig_vk_PDF_fit_DM_f50}
\end{figure}

Fig.\,\ref{Fig_vk_PDF_fit_DM_f50} shows several examples of the $v_{\rm k}$ distribution by use of the empirical $N-M_{\rm WD}$ and $\Delta M_{\rm ejct,tot}-M_{\rm WD}$ relations and  Eq.\,(\ref{Eq_vk}). Here we adopt a log-normal $f_{\rm a}$ distribution and $f_{\rm ejct}=30\%$ (left) and 50\% (right).

\newpage
\section{}\label{Sect_apdix2}
Table\,\ref{table_results} presents a list of the model parameters and the calculated results.
\begin{table*}[ht]
\centering
 \caption{Properties and predictions of multiple kick Models. 
 }
  \label{table_results}
\begin{threeparttable}

\begin{tabular}{|c|c|c|c|c|c|c|c|c|c|c|}
    \hline 
\diagbox [width=10em,trim=ll] {Name}{Properties} & \
                               $v_{\rm k}$ distribution                    & $v_{\rm k}$ limit$^{(1)}$ & $N$ & $v_{\rm k}$ in flashes & $f_{\rm a}^{(2)}$ & $e\le10^{-3}$ &  $e\ge 10^{-2}$ &  $e\ge 10^{-1}$ &  $\mu \ge 4.8^{\degree}$ &  $\mu \ge 10^{\degree}$ \\
\hline
N1\_$\sigma(+1)$\_A & Maxwell: $\sigma_{\rm k} =1^{(1)}$  &  ...                           & 1     &                        & No              &   0.2\%            &  84.0\%            &                            &                                      &   \\
\hline
N1\_$\sigma(+2)$\_A & Maxwell: $\sigma_{\rm k}$ =2  &  ...                                    & 1    &                         & No              &                      &  95.7\%            &  5.2\%                 & 0.1\%                            &           \\
\hline
N1\_$\sigma(+3)$\_A &Maxwell: $\sigma_{\rm k}$ =3  &  ...                                     & 1    &                        & No              &                        &  98.0\%           &  21.1\%              & 2.9\%                            &          \\
\hline
N3\_$\sigma(+1)$\_A & Maxwell: $\sigma_{\rm k}$ =1  &  5                                     & 3    & independent                       & No              &   0.1\%            & 94.7\%            &  1.6\%                         &                                       &           \\
\hline
N3\_$\sigma(+2)$\_A & Maxwell: $\sigma_{\rm k}$ =2  &  10                                   & 3    & independent                       & No              &                       &  98.6\%              &  29.0\%              & 4.1\%                             &           \\
\hline
N5\_$\sigma(+1)$\_A & Maxwell: $\sigma_{\rm k}$ =1  &  5                                     & 5    & independent                       & No              &                     &  96.8\%              &  5.9\%             & 0.1\%                               &          \\
\hline
N5\_$\sigma(+2)$\_A & Maxwell: $\sigma_{\rm k}$ =2  &  10                                  & 5    & independent                       & No              &                     &  99.2\%                &  44.8\%            & 111.9\%                             & 0.1\%          \\
\hline
N3\_$\sigma(-1)$\_A &Flat                                                  &  5                              & 3     & independent                       & No              &                    &  97.8\%                &  21.2\%            & 0.4\%                                &           \\
\hline
N3\_$\sigma(-2)$\_A & Flat                                                 &  10                            & 3     & independent                       & No              &                      &  99.4\%                 & 62.3\%             & 18.1\%                             & 0.3\%          \\
\hline
N5\_$\sigma(-1)$\_A & Flat                                                 &  5                             & 5     & independent                       & No              &                      &  98.6\%               &  34.9\%              & 2.4\%                                &           \\
\hline
N5\_$\sigma(-2)$\_A & Flat                                                & 10                            & 5      & independent                       & No              &                      &  99.6\%                &  72.3\%            & 34.3\%                                & 2.3\%          \\
\hline

N3\_$\sigma(+1)$\_B & Maxwell: $\sigma_{\rm k}$ =1  &  5                                 & 3      & unchanged                       & No               &   0.5\%            &  81.7\%                &  2.2\%               &                                          &           \\
\hline
N3\_$\sigma(+2)$\_B & Maxwell: $\sigma_{\rm k}$ =2  &  10                              & 3      & unchanged                       & No               &   0.1\%             &  93.3\%                &  18.3\%                & 1.4\%                                &           \\
\hline
N5\_$\sigma(+1)$\_B &Maxwell: $\sigma_{\rm k}$ =1  &  5                                 & 5      & unchanged                       & No              &   0.4\%            &  85.5\%                 &  5.5\%                 & 0.1\%                                &           \\
\hline
N5\_$\sigma(+2)$\_B & Maxwell: $\sigma_{\rm k}$ =2  &  10                              & 5    & unchanged                       & No                &   0.1\%            &  94.7\%                &  26.9\%                 & 4.4\%                                & 0.1\%          \\
\hline
N3\_$\sigma(-1)$\_B &Flat                                                  &  5                           & 3   & unchanged                      & No              &   1.2\%            &  87.5\%                &  20.3\%                 & 1.4\%                                &           \\
\hline
N3\_$\sigma(-2)$\_B & Flat                                                 &  10                         & 3   & unchanged                     & No               &   0.6\%            &  93.7\%                &  47.5\%                 & 18.8\%                                & 1.3\%          \\
\hline
N5\_$\sigma(-1)$\_B & Flat                                                 &  5                           & 5   & unchanged                       & No              &   1.1\%            &  89.4\%                &  28.5\%                  & 5.1\%                                 &           \\
\hline
N5\_$\sigma(-2)$\_B & Flat                                                & 10                           & 5   & unchanged                       & No              &   0.5\%            &  94.6\%                &  54.5\%                 & 29.3\%                                & 5.2\%        \\
\hline

N3\_$\sigma(+1)$\_B+fa & Maxwell: $\sigma_{\rm k}$ =1  &  5                          & 3   & unchanged                       & Yes             &   65.8\%            &  9.8\%                &                              &                                           &           \\
\hline
N3\_$\sigma(+2)$\_B+fa & Maxwell: $\sigma_{\rm k}$ =2  &  10                        & 3   & unchanged                       & Yes             &   58.2\%            &  16.5\%               &  0.7\%                 &                                            &           \\
\hline
N5\_$\sigma(+1)$\_B+fa & Maxwell: $\sigma_{\rm k}$ =1  &  5                          & 5   & unchanged                       & Yes             &   63.9\%            &  11.7\%                 &0.1\%                   &                                          &           \\
\hline
N5\_$\sigma(+2)$\_B+fa & Maxwell: $\sigma_{\rm k}$ =2  &  10                        & 5   & unchanged                       & Yes             &  56.4\%            &  18.6\%                &  1.2\%                 & 0.1\%                                &           \\
\hline
N3\_$\sigma(-1)$\_B+fa &Flat                                                  &  5                     & 3   & unchanged                      & Yes             &   59.6\%            &  15.9\%                &  0.8\%                  &                                        &           \\
\hline
N3\_$\sigma(-2)$\_B+fa & Flat                                                 &  10                   & 3   & unchanged                     & Yes              &   52.2\%            &  22.8\%                &  3.3\%                 & 0.6\%                                &           \\
\hline
N5\_$\sigma(-1)$\_B+fa & Flat                                                 &  5                     & 5   & unchanged                      & Yes             &   57.8\%            &  17.8\%                &  1.4\%                 & 0.1\%                                &           \\
\hline
N5\_$\sigma(-2)$\_B+fa & Flat                                                & 10                      & 5   & unchanged                      & Yes             &  50.3\%            &  24.9\%                &  4.5\%                 & 1.3\%                                & 0.1\%          \\
\hline

N5\_$\sigma(+2)$\_B+fa\_U & Maxwell: $\sigma_{\rm k}$ =2  & 10                     & 5   & unchanged                      & ``U''-like      &   31.8\%            & 51.6 \%                &  19.6\%                 & 8.4\%                              &  1.1\%          \\
\hline

N5\_$\sigma(-2)$\_B+fa\_U &  Flat                                          & 10                     & 5   & unchanged                      & ``U''-like     &  34.3\%            &  46.7\%                &  7.8\%                 & 1.0\%                                &           \\
\hline
\end{tabular}\vspace{0cm}
\begin{tablenotes}
\item[$^{\rm 1}$] in the unit of $\rm km\,s^{-1}$.
\item[$^{\rm 2}$] $f_{\rm a}$ follows a log-normal distribution unless otherwise specified.
\end{tablenotes}
\end{threeparttable}
\end{table*}

\begin{table*}
  \begin{minipage}[t]{1\textwidth}
 \caption{Properties and predictions of multiple kick Models based on the fitted $N-M_{\rm WD}$ and $\Delta M_{\rm ejct,tot}-M_{\rm WD}$ relations. 
 }
\begin{tabular}{|c|c|c|c|c|c|p{1cm}<\centering|c|c|c|c|c|}
\hline	
\diagbox [width=10em,trim=ll] {Name}{Properties} &  
                     $v_{\rm k}$ and $N$ & $Z$  & \makecell[l]{element\\diffusion}  &  $f_{\rm eject}$ &   $v_{\rm k}$ in flashes  & $f_{\rm a}$ & $e\le10^{-3}$ &  $e\ge 10^{-2}$ &  $e\ge 10^{-1}$ &  $\mu \ge 4.8^{\degree}$ &  $\mu \ge 10^{\degree}$  \\
\hline                                        
 $\rm Z02\_ED0\_f10\_B+fa$ & Fitting  & 0.02  & No                         &  10\%                &    unchanged                        &  Yes            &  59.7\%       &  15.1\%              &                         &                                          &                                 \\
\hline
 $\rm Z02\_ED0\_f30\_B+fa$ & Fitting  & 0.02  & No                         &  30\%                &    unchanged                        &  Yes           &  44.1\%        & 27.1\%              &  2.8\%               &   0.6\%                              &                                \\
\hline
$\rm Z02\_ED0\_f50\_B+fa$ & Fitting  & 0.02  & No                          &  50\%                &    unchanged                        &  Yes          &  16.6\%          & 32.7\%             &   8.0\%               &   5.3\%                              &                                 \\
\hline

$\rm Z02\_ED1\_f10\_B+fa$ & Fitting  & 0.02  & Yes                         &  10\%                &    unchanged                        &  Yes          &  61.8\%             & 13.6\%              &                       &                                         &                                 \\
\hline

$\rm Z02\_ED1\_f30\_B+fa$ & Fitting  & 0.02  & Yes                         &  30\%                &    unchanged                        &  Yes          &  48.8\%             & 25.6\%            &   4.2\%               &   0.3\%                              &                                 \\
\hline

$\rm Z02\_ED1\_f50\_B+fa$ & Fitting  & 0.02  & Yes                         &  50\%                &    unchanged                        &  Yes          &  38.6\%             & 31.2\%             &   9.3\%               &   2.2\%                              &  0.1\%                               \\
\hline
$\rm Z001\_ED0\_f10\_B+fa$ & Fitting  & 0.001  & No                         &  10\%                &    unchanged                        &  Yes          & 54.2\%         & 20.4\%                 & 0.1\%                          &                                       &                                \\
\hline
$\rm Z001\_ED0\_f30\_B+fa$ & Fitting  & 0.001  & No                         &  30\%                &    unchanged                        &  Yes          &  24.5\%          & 32.4\%                &   8.4\%              &  4.5\%                            &   0.1\%                              \\
\hline
$\rm Z001\_ED0\_f50\_B+fa$ & Fitting  & 0.001  & No                         &  50\%                &    unchanged                        &  Yes          &  7.8\%          & 38.0\%                &   14.0\%               &   9.3\%                           &   2.1\%                              \\
\hline
\end{tabular}

\end{minipage}
\end{table*}

\end{document}